\documentclass[twoside,twocolumn,9pt]{article}
\usepackage{extsizes}
\usepackage{latexsym}
\usepackage[super,sort&compress,comma]{natbib} 
\usepackage[version=3]{mhchem}
\usepackage[left=1.5cm, right=1.5cm, top=1.785cm, bottom=2.0cm]{geometry}
\usepackage{balance}
\usepackage{times,mathptmx}
\usepackage{graphicx} 
\usepackage{lastpage}
\usepackage[format=plain,justification=justified,singlelinecheck=false,font={stretch=1.125,small,sf},labelfont=bf,labelsep=space]{caption}
\usepackage{float}
\usepackage{fancyhdr}
\usepackage{fnpos}
\usepackage[english]{babel}
\usepackage[T1]{fontenc}
\usepackage{multirow}
\usepackage[bottom]{footmisc}

\usepackage{endnotes}
\usepackage{cancel}

\usepackage{amssymb,amsmath,amsfonts}
\usepackage{textcomp}
\usepackage{graphicx,color}
\usepackage[utf8x]{inputenc}
\usepackage{float}
\usepackage{subfigure}

\addto{\captionsenglish}{%
  
}
\usepackage{array}
\usepackage{droidsans}
\usepackage{charter}
\usepackage[usenames,dvipsnames]{xcolor}
\usepackage{setspace}
\usepackage[compact]{titlesec}
\usepackage{hyperref}
\usepackage{epstopdf}%This line makes .eps figures into .pdf - please comment out if not required.

\definecolor{cream}{RGB}{222,217,201}
\usepackage[normalem]{ulem} % either use this (simple) or
\usepackage{soul} % use this (many fancier options)

\renewcommand{\vec}[1]{\boldsymbol{\mathrm{#1}}}%

\begin{document}
\pagestyle{fancy}
\thispagestyle{plain}
\fancypagestyle{plain}{

%%%HEADER%%%
%\fancyhead[C]{\includegraphics[width=18.5cm]{head_foot/header_bar}}
%\fancyhead[L]{\hspace{0cm}\vspace{1.5cm}\includegraphics[height=30pt]{head_foot/SM}}
%\fancyhead[R]{\hspace{0cm}\vspace{1.7cm}\includegraphics[height=55pt]{head_foot/RSC_LOGO_CMYK}}
}
%%%END OF HEADER%%%

%%%PAGE SETUP - Please do not change any commands within this section%%%
\makeFNbottom
\makeatletter
\renewcommand\LARGE{\@setfontsize\LARGE{15pt}{17}}
\renewcommand\Large{\@setfontsize\Large{12pt}{14}}
\renewcommand\large{\@setfontsize\large{10pt}{12}}
\renewcommand\footnotesize{\@setfontsize\footnotesize{7pt}{10}}
\makeatother

\makeatletter 
\renewcommand\@biblabel[1]{#1}            
\renewcommand\@makefntext[1]% 
{\noindent\makebox[0pt][r]{\@thefnmark\,}#1}
\makeatother 
\renewcommand{\figurename}{\small{Fig.}~}
%\sectionfont{\sffamily\Large}
%\subsectionfont{\normalsize}
%\subsubsectionfont{\bf}
\setstretch{1.125} %In particular, please do not alter this line.
\setlength{\skip\footins}{0.8cm}
\setlength{\footnotesep}{0.25cm}
\setlength{\jot}{10pt}
\titlespacing*{\section}{0pt}{4pt}{4pt}
\titlespacing*{\subsection}{0pt}{15pt}{1pt}
%%%END OF PAGE SETUP%%%

%%%FOOTER%%%
\fancyfoot{}
\fancyfoot[LO,RE]{\vspace{-7.1pt}\includegraphics[height=9pt]{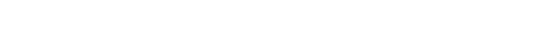}}
\fancyfoot[CO]{\vspace{-7.1pt}\hspace{13.2cm}\includegraphics{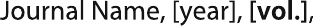}}
\fancyfoot[CE]{\vspace{-7.2pt}\hspace{-14.2cm}\includegraphics{head_foot/RF}}
\fancyfoot[RO]{\footnotesize{\sffamily{1--\pageref{LastPage} ~\textbar  \hspace{2pt}\thepage}}}
\fancyfoot[LE]{\footnotesize{\sffamily{\thepage~\textbar\hspace{3.45cm} 1--\pageref{LastPage}}}}
\fancyhead{}
\renewcommand{\headrulewidth}{0pt} 
\renewcommand{\footrulewidth}{0pt}
\setlength{\arrayrulewidth}{1pt}
\setlength{\columnsep}{6.5mm}
\setlength\bibsep{1pt}
%%%END OF FOOTER%%%

%%%FIGURE SETUP - please do not change any commands within this section%%%
\makeatletter 
\newlength{\figrulesep} 
\setlength{\figrulesep}{0.5\textfloatsep} 

%\newcommand{\topfigrule}{\vspace*{-1pt}% 
%\noindent{\color{cream}\rule[-\figrulesep]{\columnwidth}{1.5pt}} }

%\newcommand{\botfigrule}{\vspace*{-2pt}% 
%\noindent{\color{cream}\rule[\figrulesep]{\columnwidth}{1.5pt}} }

%\newcommand{\dblfigrule}{\vspace*{-1pt}% 
%\noindent{\color{cream}\rule[-\figrulesep]{\textwidth}{1.5pt}} }

%\makeatother
%%%END OF FIGURE SETUP%%%
%%%TITLE, AUTHORS AND ABSTRACT%%%
\twocolumn[
  \begin{@twocolumnfalse}
\vspace{3cm}
\sffamily
\begin{tabular}{m{4.5cm} p{13.5cm} }

\includegraphics{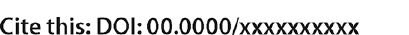} & \noindent\LARGE{\textbf{Rheology of active polar emulsions: from linear to unidirectional and unviscid flow, and intermittent viscosity$^\dag$}} \\%Article title goes here instead of the text "This is the title"
%\vspace{0.3cm} & \vspace{0.3cm} \\

 &\noindent\large{G. Negro$^{\ast}$\textit{$^{a}$}, L.N. Carenza$^{\ast}$\textit{$^{a}$},  A. Lamura\textit{$^{b}$}, A. Tiribocchi$^{\ast}$\textit{$^{c}$} and G. Gonnella\textit{$^{a\ddag}$}} \\%Author names go here instead of "Full name", etc.

\includegraphics{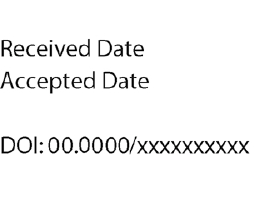} & \noindent\normalsize{The rheological behaviour of an emulsion made of an active polar  component and an  isotropic passive fluid  is studied by lattice Boltzmann methods. Different flow regimes are found by varying the values of shear rate and extensile activity~(occurring, e.g., in microtubule-motor suspensions). By increasing activity,  a first transition occurs from  linear flow  regime to spontaneous persistent unidirectional macro-scale flow,  followed by another  transition either to (low shear) intermittent  flow regime  with    coexistence of states  with positive, negative, and vanishing apparent viscosity,
or to (high shear) symmetric shear thinning regime.
The different behaviours can be explained in terms of the dynamics of the  polarization field close to the walls. A maximum entropy production principle selects the most likely states in the intermittent regime.
} \\%The abstrast goes here instead of the text "The abstract should be..."

\end{tabular}

 \end{@twocolumnfalse} \vspace{0.6cm}

  ]

%%%END OF TITLE, AUTHORS AND ABSTRACT%%%

%%%FONT SETUP - please do not change any commands within this section
\renewcommand*\rmdefault{bch}\normalfont\upshape
\rmfamily
\section*{}
\vspace{-1cm}

%%%FOOTNOTES%%%

\footnotetext{\textit{$^{a}$~Dipartimento  di  Fisica,  Universit\`a  degli  Studi  di  Bari  and  INFN,
Sezione  di  Bari,  via  Amendola  173,  Bari,  I-70126,  Italy}}
\footnotetext{\textit{$^{b}$~Istituto Applicazioni Calcolo, CNR, Via Amendola 122/D, I-70126 Bari, Italy}}
\footnotetext{\textit{$^{c}$~Center for Life Nano Science@La Sapienza, Istituto Italiano di Tecnologia, 00161 Rome, Italy and Istituto per le Applicazioni del Calcolo CNR, via dei Taurini 19, 00185 Rome, Italy
}}

%Please use \dag to cite the ESI in the main text of the article.
%If you article does not have ESI please remove the the \dag symbol from the title and the footnotetext below.
%\footnotetext{\dag~Electronic Supplementary Information (ESI) available: [details of any supplementary information available should be included here]. See DOI: 10.1039/cXsm00000x/}
%additional addresses can be cited as above using the lower-case letters, c, d, e... If all authors are from the same address, no letter is required

%\footnotetext{\ddag~Additional footnotes to the title and authors can be included \textit{e.g.}\ `Present address:' or `These authors contributed equally to this work' as above using the symbols: \ddag, \textsection, and \P. Please place the appropriate symbol next to the author's name and include a \texttt{\textbackslash footnotetext} entry in the the correct place in the list.}

\balance

%%%MAIN TEXT%%%%
\section{Introduction}
Active gels~\cite{kruse2004,joanny2009,marchetti2013} are a new class of complex fluids with striking physical properties and many possible innovative applications~\cite{Sokolov2009,DiLeonardo2010,Yeomans2014,doostmohammadi2018,needleman2017}. As other kinds of active systems~\cite{vicsek2012, poon2013, elgeti2015, cates2015, kanso2015, bechinger2016, yeomans2017, gonnella2015, digregorio2018}, they  are maintained in their driven state -- far from thermodynamic equilibrium -- by  energy  supplied directly and independently at the level of individual constituents. Examples are  suspensions of biological filaments, such as actomyosin and microtubule bundles, activated  with  motor proteins~\cite{surrey2001,prost2015,guillamat2016,sanchez2012} and bacterial cultures~\cite{dombrowski2004,zhang2010}. The  constituents of these  systems   have  the natural tendency to assemble and align, thus developing structures with typical polar or nematic order. Combination of this property with self-motility capacity  is at the origin of a wealth  of interesting phenomena,  not observable in absence of  activity  \cite{ramaswamy2010}, including spontaneous flow~\cite{simha2002,kruse2004,marenduzzo2007},  active turbulence at low Reynolds numbers~\cite{dombrowski2004,wensink2012,carenza19turb}, and unusual rheological  properties~\cite{marchetti2013,saintillan2018}. Most of these behaviours were found in single component fluids, while mixtures of active and passive components have not been too much investigated so far~\cite{cates2018}.

Complex  rheological behaviours in active matter depend on the  interplay  between the external forcing and the circulating flow induced by active agents. For instance, the swimming mechanism of pusher microswimmers, like \textit{E. Coli}, produces a far flow field characterized by quadrupolar symmetry, in which fluid is expelled along the fore-aft axes of the swimmer and drawn trasversely, thus leading to \textit{extensile} flow patterns. These enforce the applied flow, in the case of flow-aligning  swimmers, causing  shear thinning~\cite{hatwalne2004}. This may lead to the  occurrence of a \textit{superfluidic} regime  with vanishing  (apparent) shear viscosity that was speculated in~\cite{cates2008} for the case of an active liquid crystals close to the isotropic-nematic transition. Experiments~\cite{sokolov2009_2,gachelin2013} and further theories~\cite{giomi2010} confirmed that extensile active components are able to lower the viscosity of thin film suspensions. An effective inviscid flow  was observed in~\cite{lopez2015} and more recently in~\cite{guo2018}, when the concentration and activity  of \textit{E. Coli} are sufficiently large to support coherent collective swimming. A related feature in extensile gels is the appearance of persistent uni-directional flows in  experiments on bacterial suspensions~\cite{wioland2016} and ATP-driven gels~\cite{wu2017}. Recently, numerical simulations in quasi-$1d$ geometries and linear analysis of active polar liquid crystal models have shown the occurrence of vanishing and even negative viscosity  states~\cite{loisy2018}. However, a complete characterization of the rheology of fully $2d$ active compounds has never been accomplished so far, despite being fundamental to unveil dynamical mechanisms leading to the complex properties  presented. We also mention that puller swimmers -- exerting a contractile force dipole on the surrounding fluid -- still generate a quadrupolar far flow field, but this time the fluid is expelled trasversely to its body. This explains the shear thickening behaviour, observed in experiments performed on suspensions of \textit{C. reinhardtii}~\cite{Rafai2010} -- a species of micro-alga that propels itself by means of two flagella producing contractile movements -- and in numerical studies on contractile gels~\cite{marenduzzo2007,giomi2010,foffano2012}. In this paper we will focus on extensile systems.

We study, by extensive lattice Boltzmann simulations,  the rheology of a $2d$ emulsion, made of an active fluid component (a polar gel) and an isotropic passive fluid, under simple shear flows. The system that we consider \cite{bonelli2019,negro2018} has the property that a  tunable amount of active material  can be homogeneously dispersed in an emulsion. With respect to the single-component active gel theory, this approach has the great advantage that the typical length-scale of active injection can be kept under control by  setting opportunely the parameters of the model. This would also delucidate the mechanisms for the rich rheological behaviour previously described. Experimental realization can be achieved by either confining activated cellular extracts in an emulsion of water-in-oil~\cite{sanchez2012, guillamat2016} or dispersing bacteria in water, ensuring microphase separation through depletion forces and a suitable surfactant~\cite{SchwarzLinek2012}. 

In Sect. 2 we describe the model for active emulsions and the lattice Boltzmann method implemented for the  numerical study. Moreover, we  introduce our main adimensional parameters that are related to the strength of both active injection and shear flow. In Sect. 3 we describe the numerical results for the flow behaviour in a region of the parameter space characterized by a double transition, first from  linear velocity profiles to unidirected motion and then, by increasing activity,  to symmetric shear thinning behaviour. We describe the relevance of the role of the polarization and active stress in proximity of external walls. In Sect. 4 intermittent flow -- occurring in a region of the parameter space at low shear rate and sufficiently high viscosity --  will be analyzed, also in terms of the entropy production rate evaluated for the different coexisting flow states. Finally, in Sect. 5 we  show the phase diagram summarizing results for the different behaviours we have found and we draw some conclusions.

\section{Model and Numerical Methods}
We outline here the hydrodynamic model and the numerical method used to conduct our study. We consider a fluid in $2d$ comprising a mixture of active material and an isotropic solvent with total mass density $\rho$.
The physics of the resulting composite material can be described by using an extended version of the well-established active gel theory~\cite{ramaswamy2010,marchetti2013,tjhung2012,tjhung2015,elsen,kruse2004}. The hydrodynamic variables are the density of the fluid $\rho$, its velocity ${\bf v}$, the  concentration of the active material $\phi$, and the polarization ${\bf P}$, which accounts for the average orientation of the active constituents. The dynamical equations ruling the evolution of the system in the incompressible limit are:
\begin{eqnarray}
\rho\left(\frac{\partial}{\partial t}+\mathbf{v}\cdot\nabla\right)\mathbf{v} & = &  \nabla\cdot \tilde{\sigma}^{tot}\ ,\label{nav}\\
\frac{\partial \phi}{\partial t}+\nabla\cdot\left(\phi\mathbf{v}\right)&=& M \nabla^2  \mu,\label{conc_eq}\\
\frac{\partial\mathbf{P}}{\partial t}+\left(\mathbf{v}\cdot\nabla\right)\mathbf{P}&=&-\tilde{\Omega}\cdot\mathbf{P}+\xi\tilde{D}\cdot\mathbf{P}
-\frac{1}{\Gamma} \mathbf{h}.\label{P_eq}
\end{eqnarray}
The first one is the Navier-Stokes equation, where $\tilde{\sigma}^{tot}$ is the total stress tensor \cite{beris1994}.
Eqs.~\eqref{conc_eq}-\eqref{P_eq} govern the time evolution of the concentration of the active material and of the polarization field, respectively. By the assumption that the amount of the active component is locally conserved, the evolution of the concentration field, Eq.~\eqref{conc_eq},  can be written  as a convection-diffusion equation, where $M$ is the mobility, $\mu=\delta F/\delta\phi$ the chemical potential, with $F$ a suitable free energy functional, encoding the equilibrium properties of the system, to be defined later. The dynamics of the polarization field follows an advection-relaxation equation, Eq.~\eqref{P_eq}, borrowed from polar liquid crystal theory. Here $\Gamma$ is the rotational viscosity, $\xi$ is a constant controlling the aspect ratio of active particles (positive for rod-like particles and negative for disk-like ones) and their response to external flows ($|\xi|>1$ for flow-aligning particles and $|\xi|<1$ for flow-tumbling ones), while $\mathbf{h}=\delta F/\delta\mathbf{P}$ is the molecular field. $\tilde{D}=(\tilde{W}+\tilde{W}^T)/2$ and $\tilde{\Omega}=(\tilde{W}-\tilde{W}^T)/2$ respectively represent the symmetric and the anti-symmetric parts of the velocity gradient tensor $W_{\alpha\beta}=\partial_{\beta}v_{\alpha}$, where Greek indexes denote Cartesian components. These contributions are in addition to the material derivative, as the liquid crystal can be rotated or aligned by the fluid~\cite{beris1994}.

The stress tensor $\tilde{\sigma}^{tot}$ considered in the Navier-Stokes equation, Eq.~(\ref{nav}), can be splitted into an equilibrium/passive and a non-equilibrium/active part:
\begin{equation}
\tilde{\sigma}^{tot}=\tilde{\sigma}^{pass}+\tilde{\sigma}^{act} \mbox{.}
\end{equation}
The passive term accounts for the viscous dissipation and reactive phenomena -- the elastic response of the binary fluid and of the liquid crystal --  and is given by four contributions:
\begin{equation}
\tilde{\sigma}^{\textit{pass}}=\tilde{\sigma}^{\textit{hydro}}+\tilde{\sigma}^{\textit{visc}}+\tilde{\sigma}^{\textit{pol}}+\tilde{\sigma}^{bm} \mbox{.}
\label{eqn:passive_stress_tensor}
\end{equation}
The first term is the hydrodynamic pressure contribution given by $\sigma^{\textit{hydro}}_{\alpha\beta}=-p\delta_{\alpha\beta}$. The second term is the viscous stress, written as  $\sigma_{\alpha\beta}^{visc}=\eta_0(\partial_{\alpha}v_{\beta}+\partial_{\beta}v_{\alpha})$, where $\eta_0$ is the shear viscosity. The second term is the polar elastic stress, analogous to the one used in nematic liquid crystal hydrodynamics~\cite{beris1994}:
\begin{eqnarray}
\sigma_{\alpha\beta}^{pol}=\frac{1}{2}(P_{\alpha}h_{\beta} -P_{\beta}h_{\alpha})-\frac{\xi}{2}(P_{\alpha}h_{\beta}+P_{\beta}h_{\alpha})\nonumber\\
- k_{P} \partial_{\alpha}P_{\gamma}\partial_{\beta}P_{\gamma}\label{eq:elastic-stress},
\end{eqnarray}
where $k_P$ is the elastic constant of the liquid crystal, in the single constant approximation~\cite{degennes1993}. The third term on the right-hand side of Eq.~\eqref{eqn:passive_stress_tensor} is borrowed from binary mixture theories and it includes interfacial contribution between the two phases:
\begin{equation}
\sigma_{\alpha\beta}^{bm}=\left( f-\phi\frac{\delta F}{\delta\phi} \right)\delta_{\alpha\beta} - \frac{\delta F}{\partial\left(\partial_{\beta}\phi\right)} \partial_{\alpha}\phi,
\end{equation}
where we have denoted with $f$ the free energy density. The active contribution to the stress tensor, the only one not stemming from the free energy, is given by~\cite{elsen,simha2002}
\begin{equation}
\sigma_{\alpha\beta}^{act}=-\zeta \phi \left(P_{\alpha}P_{\beta}-\frac{1}{3}|{\bf P}|^2\delta_{\alpha\beta}\right)\label{eq:active-stress},
\end{equation}
where $\zeta$ is the activity strength, positive for extensile systems (pushers) and negative for contractile ones (pullers). The active stress drives the system out of equilibrium, by injecting energy on the typical lengthscales of deformation of the polarization pattern.
\begin{figure*}[t]
\centering
{\includegraphics[width=.96\textwidth]{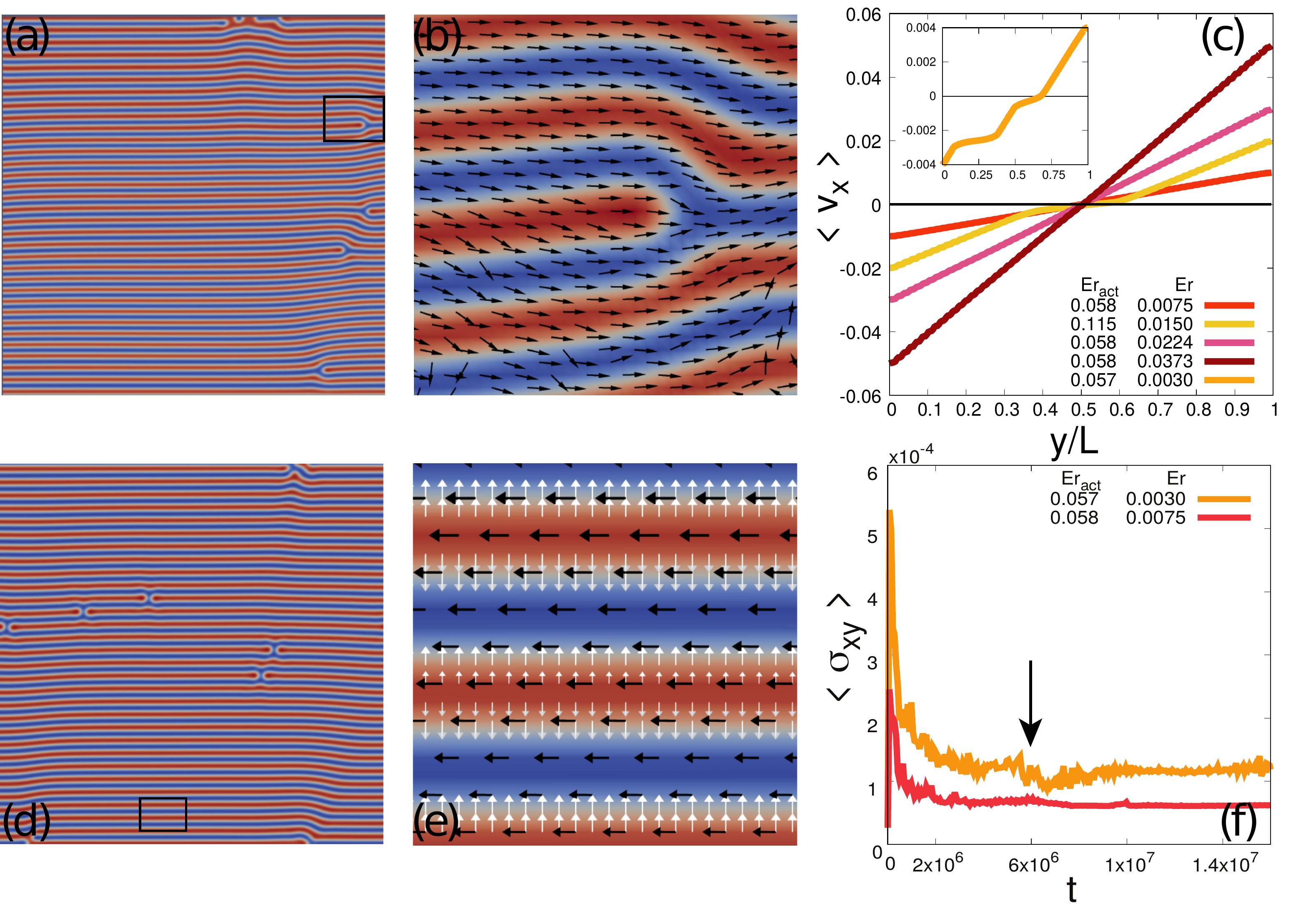}}
\caption{\textit{Linear velocity profiles and lamellar phase.} Concentration contour plots at
$Er_{act}=0.057$, $Er=0.0030$ and $Er_{act}=0.058$, $Er=0.0075$ are respectively shown in panels (a) and (d).
Here (and in the rest of the paper) red regions correspond to the active phase $(\phi \approx 2.0)$, while blue regions to the passive fluid $(\phi \approx 0.0)$.
Panels (b) and (e) show a zoom of the regions highlighted by the black squares in panels (a) and (d). Black and white vectors respectively denote velocity and polarization fields in panel (e), while only the velocity field has been plotted in panel (b). Panels (c) and (f) show the averaged velocity profile and the time evolution of the shear stress (red and orange curves correspond to the cases considered in panel (a) and (d)). Black arrow in panel~(f) points to a jump in the relaxation of $\sigma_{xy}$, due to the annihilation of two dislocations. Note that in the present Figure, as well as in the followings, the colors of the velocity and the stress profiles have been chosen in relation to the colors of the corresponding region of the phase diagram of Fig.~\ref{fig:phase_diagram}. In panels~(c-f) the profiles have been plotted in such a way to make reference to the red/orange region of linear profiles.}
\label{fig:S1}
\end{figure*}

By assuming that local equilibrium is satisfied even in presence of activity, the thermodynamics forces in Eq.~\eqref{nav}-\eqref{P_eq} can be deduced
by the following free-energy functional, coupling 
the Landau-Brazovskii model~\cite{braz,gonnella1997} to the distortion free-energy of a polar system:
%% are be encoded in
%The thermodynamics properties of the polar emulsion \textcolor{red}{that in absence of activity and under the assumption of local equilibrium~\cite{degroot1962}. These} are encoded in the following free-energy functional that couples the Landau-Brazovskii model~\cite{braz,gonnella1997} to the distortion free-energy of a polar system:
\begin{eqnarray}\label{eqn:fe}
&F[\phi,\mathbf{P}]
=\int d\mathbf{r}\,\{\frac{\tilde{a}}{4\phi_{cr}^4}\phi^{2}(\phi-\phi_0)^2+\frac{k_\phi}{2}\left|\nabla \phi\right|^{2}+\frac{c}{2}(\nabla^2\phi)^2 \nonumber\\
&-\frac{\alpha}{2} \frac{(\phi-\phi_{cr})}{\phi_{cr}}\left|\mathbf{P}\right|^2+ \frac{\alpha}{4}\left|\mathbf{P}\right|^{4}+\frac{k_P}{2}(\nabla\mathbf{P})^{2}
+\beta\mathbf{P}\cdot\nabla\phi\} \ \ .
\end{eqnarray}
This is a generalization of the free energy functional for active binary mixtures defined in~\cite{elsen}. The first term allows for the segregation of the two phases when the bulk energy density $a>0$, so that free energy has two minima at $\phi=0,\phi_0$. The second and third terms determine the interfacial tension. Notice that here a negative value of $k_\phi$ favours the formation of interfaces while a positive value of $c$ is used to guarantee thermodynamic stability~\cite{braz}. Lowering $k_\phi$ from positive to negative values leads the system to move from pure ferromagnetic phase to configurations where interfaces between components are favoured~\cite{Patzold1996,patzold1996_2,Gompper_critic,gompper1993,jaju2016}. In the Appendix \emph{Adimensional Numbers} we will show that, for a symmetric composition of the mixture, the system sets into a lamellar phase modulated at wavenumber $\kappa=\sqrt{|k_\phi|/2c}$, when $a<k^2_\phi/4c + \beta^2/k_P$ (see panels~(a) and~(b) of Fig.~\ref{fig:S1} for typical lamellar morphology).

The bulk properties of the polar liquid crystal are instead controlled by the $|\textbf{P}|^2$ and $|\mathbf{P}|^{4}$ terms, multiplied by the positive constant $\alpha$. The choice of $\phi_{cr}$ has been made in order to break the symmetry between the two phases and confine the polarization field in the active phase, $\phi > \phi_{cr}$. The term proportional to $(\nabla\mathbf{P})^{2}$ describes the energy cost due to elastic deformation in the liquid crystalline phase, gauged to the theory through the elastic constant $k_P$ (a more general treatment can be found in~\cite{reviewepje}). Finally, the last term takes into account the orientation of the polarization at the interface of the fluid. If $\beta\ne 0$, $\textbf{P}$ preferentially points perpendicularly to the interface (normal anchoring): towards the passive (active) phase if $\beta>0$ ($\beta<0$). A plot of the typical polarization field configuration to interfaces in the lamellar phase is shown with white arrows in panel~(e) of Fig.~\ref{fig:S1}.
This allows for the confinement of the active behaviour on small scales in a low-Reynolds number environment, offering a way to control the typical length-scales of energy injection -- an actual challenge for experiments in the field~\cite{sanchez2012}.
The model exhibits a wide phenomenology at varying the activity parameter $\zeta$~\cite{bonelli2019}, that has been summarized at the beginning of Sect.~\ref{sec:linear}.

\subsection{Numerical Method and Parameters}
\begin{figure*}[t]
\centering
{\includegraphics[width=1.0\textwidth]{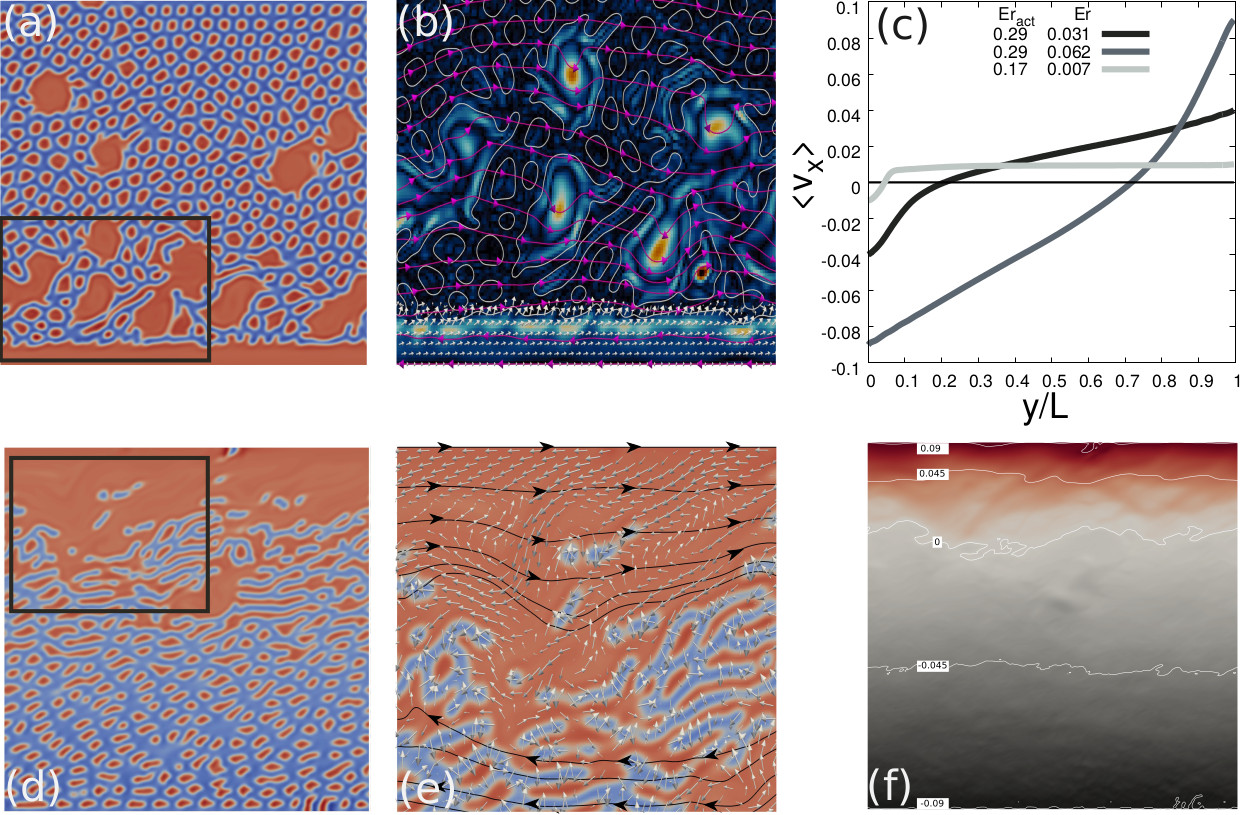}}
\caption{\textit{Unidirectional motion}. (a) Concentration contour plot at $Er_{act}=0.29$, $Er=0.031$. (b) Contour plot of the vorticity $\omega=(\partial_y v_x - \partial_x v_y)/2$ (black corresponds to $\omega=0$ and red to $\omega=10^{-2}$) in the region framed with black box in panel (a); purple lines represent velocity streamlines while the polarization field is plotted in white (for graphical clarity, only in the bottom active layer). (d) Contour plot of the concentration field at $Er_{act}=0.29$ and $Er=0.062$ and (e) zoom of the region framed with black box in panel (d), with black lines representing $\mathbf{v}$ streamlines and $\mathbf{P}$ plotted in white. (f) Contour plot of the velocity field in the flow direction for the same case shown in panel (d), with few isolines plotted in white. Panel (c) shows $\langle v_x \rangle$ for the two cases considered and also for the case at $Er_{act}=0.172$, $Er=0.0075$.}
\label{fig:S3}
\end{figure*}
The equations of motion of the polar active emulsion, Eqs.~\eqref{nav}-\eqref{P_eq},  are solved by means of a hybrid lattice Boltzmann (LB) scheme, which combines a predictor-corrector LB treatment for the Navier-Stokes equation~\cite{denniston2001}  with a finite-difference predictor-corrector algorithm to solve the order parameters dynamics, implementing a first-order upwind scheme, for the convection term, and fourth-order accurate stencil for the computation of space derivatives.

In this approach the evolution of the fluid is described in terms of a set of distribution functions ${f_i(\mathbf{r}_\alpha,t)}$ (with index $i$ labelling different lattice directions, thus ranging from $1$ to $9$) defined on each lattice site $\mathbf{r}_\alpha$. Their evolution follows a discretized predictor-corrector version of the Boltzmann equation in the BGK approximation:
\begin{eqnarray}
f_i (\mathbf{r}_\alpha + \vec{\xi}_i \Delta t) - f_i (\mathbf{r}_\alpha,t) = \nonumber \\ - \dfrac{\Delta t }{2} \left[ \mathcal{C}(f_i,\mathbf{r}_\alpha, t) + \mathcal{C}(f_i^*,\mathbf{r}_\alpha+ \vec{\xi}_i \Delta t, t) \right].
\label{eqn:LBevolution}
\end{eqnarray}
Here $\lbrace \vec{\xi}_i \rbrace$ is the set of discrete velocities, that for the $d2Q9$  model are $\vec{\xi}_0=(0,0)$, $\vec{\xi}_{1,2}=(\pm u,0)$, $\vec{\xi}_{3,4}=(0,\pm u)$, $\vec{\xi}_{5,6}=(\pm u, \pm u)$, $\vec{\xi}_{7,8}=(\pm u,\pm u)$, where $u$ is the lattice speed. The distribution functions $f^*$ are first-order estimations to  $f_i (\mathbf{r}_\alpha + \vec{\xi}_i \Delta t) $ obtained by setting $f_i^* \equiv f_i$ in Eq.~\eqref{eqn:LBevolution}, and $\mathcal{C}(f,\mathbf{r}_\alpha, t)=-(f_i-f_i^{eq})/\tau + F_i$ is the collisional operator in the BGK approximation expressed in terms of the equilibrium distribution functions $f_i^{eq}$ and supplemented with an extra forcing term for the treatment of the anti-symmetric part of the stress tensor. The density and momentum of the fluid are defined in terms of the distribution functions as follows:
\begin{equation}
\sum_i f_i = \rho \qquad \sum_i f_i \vec{\xi}_i = \rho \mathbf{v}.
\label{eqn:variables_hydro}
\end{equation}
The same relations also hold for the equilibrium distribution functions, thus ensuring mass and momentum conservation. In order to correctly reproduce the Navier-Stokes equation we impose the following condition on the second moment of the equilibrium distribution functions:
\begin{equation}
\sum_i f_i \vec{\xi}_i \otimes \vec{\xi}_i = \rho \mathbf{v} \otimes \mathbf{v} -\tilde{\sigma}^{bin} - \tilde{\sigma}^{pol}_s,
\label{eqn:constrain_second_moment}
\end{equation}
and on the force term:
\begin{align}
&\sum_i F_i = 0, \nonumber \\ &\sum_i F_i \vec{\xi}_i = \mathbf{\nabla} \cdot (\tilde{\sigma}^{pol}_a + \tilde{\sigma}^{act}), \\  &\sum_i F_i \vec{\xi}_i \otimes \vec{\xi}_i = 0, \nonumber
\label{eqn:constraint_force}
\end{align}
where we denoted with $\tilde{\sigma}^{pol}_s$ and $\tilde{\sigma}^{pol}_a$ the symmetric and anti-symemtric part of the polar stress tensor, respectively. The equilibrium distribution functions are expanded up to the second order in the velocities:
\begin{equation}
f_i^{eq} = A_i + B_i (\vec{\xi} \cdot \mathbf{v}) + C_i |\mathbf{v} |^2 + D_i (\vec{\xi} \cdot \mathbf{v})^2 + \tilde{G}_i : (\vec{\xi} \otimes \vec{\xi}).
\end{equation}
Here coefficients $A_i, B_i,C_i,D_i,\tilde{G}_i$ are to be determined imposing conditions in Eqs.~\eqref{eqn:variables_hydro} and \eqref{eqn:constrain_second_moment}. In the continuum limit the Navier-Stokes equation is restored if $\eta_0=\tau/3$.

We made use of a parallel approach implementing Message Passage Interface (MPI) to parallelize the code. We divided the computational domains in slices, and assigned each of them to a particular task in the MPI communicator. Non-local operations (such as derivatives), have been treated through the \emph{ghost-cell} approach~\cite{MPI}.

Simulations have been performed on square lattice of size $L=256$. The concentration $\phi$ ranges from $\phi\simeq 0$ (passive phase) to $\phi\simeq 2$ (active phase). Unless otherwise stated, parameter values are $a=4\times 10^{-3}$, $k_\phi=-6\times 10^{-3}$, $c=10^{-2}$, $\alpha=10^{-3}$, $k_P=10^{-2}$, $\beta=0.01$, $\Gamma=1$, $\xi=1.1$, $\phi_0=2.0$, and $\eta_0=1.67$. All quantities in the text are reported in lattice units. We initialized the system starting from a uniform phase, with $\phi (\mathbf{r})= \langle \phi \rangle + \delta \phi(\mathbf{r})$, where $\langle \phi \rangle = \phi_{cr}$ is the conserved (area) averaged value of the concentration field and $\delta \phi$ is a small perturbation field favouring phase separation and ranging in $\left[ -\frac{\phi_{cr}}{10}, \frac{\phi_{cr}}{10} \right]$. The initial condition for the polarization field is completely random, being its orientation randomly distributed in the plain, while its intensity is randomly chosen in $\left[0, 1 \right]$.

Our choice of parameters is such that the Schmidt number ($Sc = \frac{\eta_0}{\rho D} $ where  $D= 2 M a/\phi^4_{cr}$ is the diffusion constant) is fixed at values typical for liquids, $ \sim 2000$ where,
in absence of activity, lamellae show low resistance to the flow and  can easily order. It was shown in~\cite{jaju2016} that at smaller $Sc$ lamellar domains hardly align to the flow and may eventually undergo pearling instability, persistent even in the long dynamics.

We considered flow in a channel with no-slip boundary conditions at the top and the bottom walls ($y=0$ and $y=L$), implemented by bounce-back boundary conditions for the distribution functions~\cite{succi2001}, and periodic boundary conditions in the $y$ direction. The flow is driven by moving walls, respectively with velocity $v_w$ for the top wall and $-v_w$ for the bottom wall, so that the shear rate is given by $\dot{\gamma}=\frac{2 v_w}{L}$.

Moreover neutral wetting boundary conditions were enforced by requiring on the wall sites that the following relations hold:
\begin{align}
\nabla_{\perp} \mu =  0 \ , \ \ \  
\nabla_{\perp} (\nabla^2 \phi)=  0 \ ,
\end{align}
where $\nabla_{\perp}$ denotes the partial derivative computed normally to the walls and directed towards the bulk of the system.
Here the first condition ensures density conservation, the second determines the wetting to be neutral. In the case of bacterial swimmers, it is commonly observed that, close to the boundaries, they orient along the wall direction~\cite{PhysRevE.84.041932}. In  actomyosin solutions, the actin filaments can also be assumed to be anchored parallel to the walls due to focal adhesion~\cite{wozniak2004}. Therefore, suitable boundary conditions for the polarization $\mathbf{P}$ is a strong anchoring condition with $\mathbf{P}$ aligned parallel to the walls
\begin{equation}
P_\perp \rvert_\textrm{walls} =  0, \qquad \nabla_\perp P_\parallel \rvert_\textrm{walls}=  0 ,
\end{equation}
where $P_\perp$ and $P_\parallel$ denote, respectively, normal and tangential components of the polarization field with respect to the walls. In order to compare external and active forcing in our system, we make use of the Ericksen number, $Er$, and the \emph{active} Ericksen number $Er_{act}$ as relevant adimensional quantities. The former is often used in the study of liquid crystals to describe the deformation of the orientational order parameter field under flow and it is defined as the ratio of the viscous stress to the elastic stress. In particular in lamellar systems a suitable choice is given by:
\begin{equation}
\textrm{Er}=\frac{\eta_0\dot\gamma}{B}\ ,
\end{equation}
where $B$ is the lamellar compression modulus -- namely the energy cost for the variation of the lamellar width $\lambda=2\pi/\kappa$ per unit length, whose expression in terms of the parameter of the model is explicitly derived in the Appendix~A. The active Ericksen number, suggested by Giomi for the first time in~\cite{Giomi20130365}, is, in turn, defined as the ratio between the module of the activity parameter $\zeta$ and the compression modulus:
\begin{equation}
\textrm{Er}_{act}=\frac{|\zeta|}{B}.
\end{equation}

\begin{figure}
\centering
{\includegraphics[width=.48\textwidth]{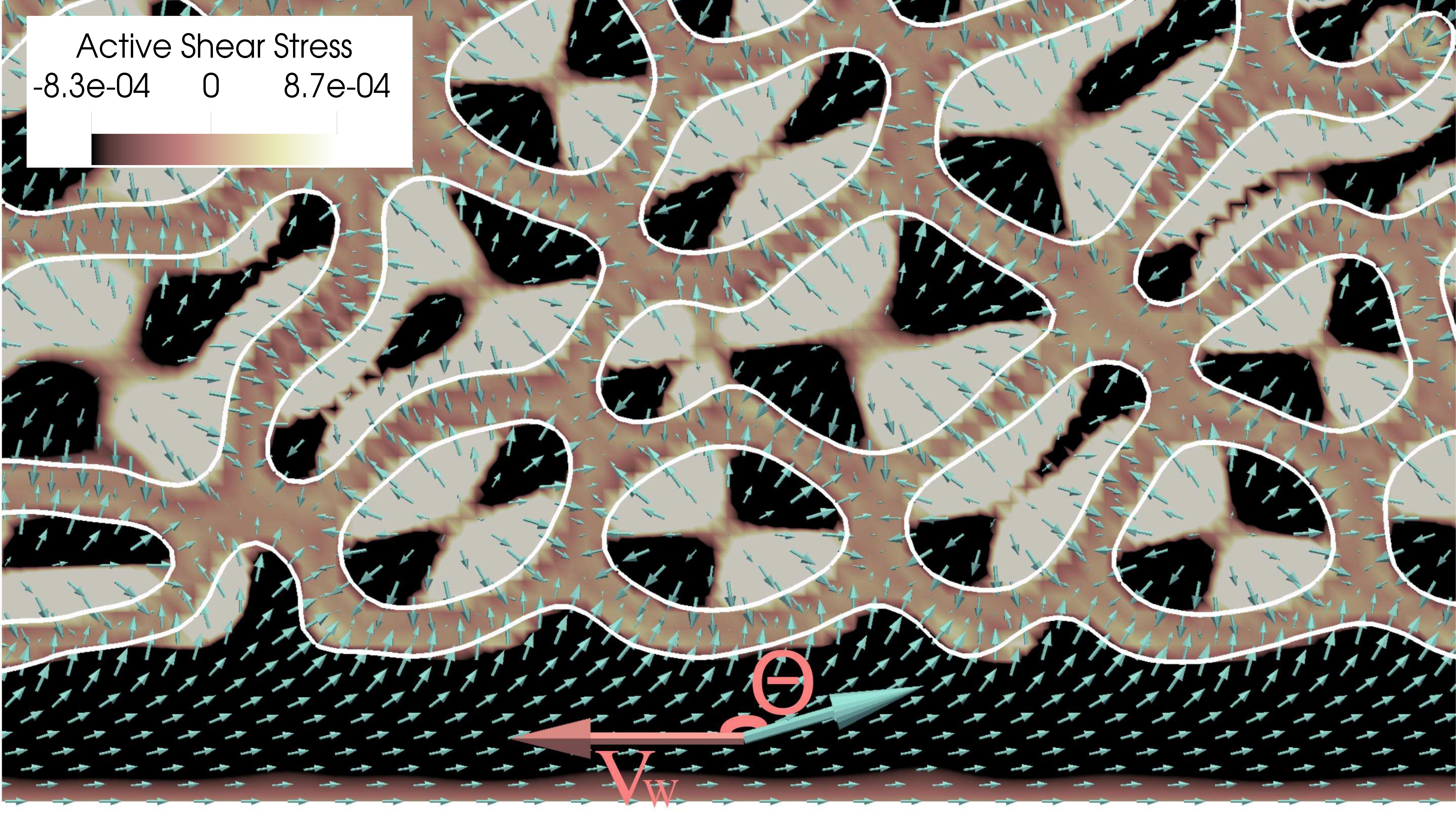}}
\caption{
Active shear stress close to a wall at $Er=0.0074$, $Er_{act}=0.172$.
The polarization field $\mathbf{P}$ (cyan arrows) exhibits a splayed profile under the mutual effect of strong anchoring to the wall (tangential) and to the interface (homeotropic), while the red arrow shows the direction of the imposed velocity $\mathrm{v_w}$.
The angle $\theta$ denotes the local orientation of the polarization, sketched by the magnified reference cyan arrow, with respect to the flow. White lines trace the interface ($\phi=\phi_{cr}$) between active and passive phases.
Passive (beige) regions are almost stress free, while  negative stress in the boundary layer (black) corresponds to a net force opposite to the flow direction. Droplets have quadrupolar structures.
}
\label{fig:pol_at_walls}
\end{figure}

\section{Linear flow and symmetry breaking transition}
\label{sec:linear}

\begin{figure*}[t]
\centering
{\includegraphics[width=.98\textwidth]{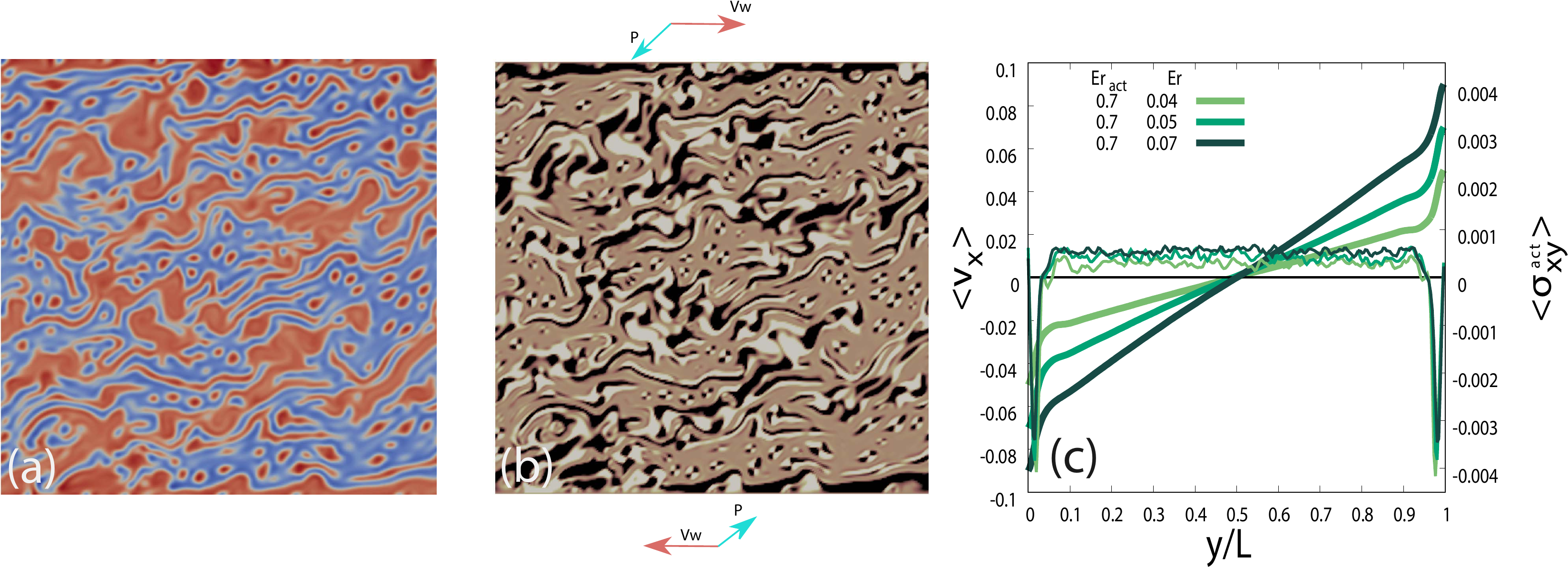}}
\caption{\textit{Symmetric shear thinning profiles}.
Concentration (a) and active shear stress (b) contour plot for the case at $Er_{act}=0.70 , Er=0.04$. The colour code is the same of Fig.\ref{fig:pol_at_walls}.
Red arrows denote the direction of the moving walls, and cyan ones the direction of the polarization, averaged within the active layers close to the walls.
(c) Velocity $\langle v_x \rangle $ (thick lines) and active shear stress profiles $\langle \sigma_{xy}^{act} \rangle $ (thin) averaged along the flow direction, for the case corresponding to panels (a) and (b) and analogous ones.}
\label{fig:S4}
\end{figure*}

Before presenting specific results case we summarize the morphological phenomenology arising in an active extensile polar lamellar system at varying the activity parameter ($\zeta>0$) in absence of any external forcing. Bonelli \emph{et al.}~\cite{bonelli2019} showed that the shear-free system is characterized by  a transition at $ Er_{act} \approx 0.11 $  from the lamellar phase to an emulsion with moving active droplets. The bending instability, typical of extensile gels~\cite{simha2002}, favors this rearrangement. For $ Er_{act} \gtrsim 1 $  the system enters in a totally mixed phase, characterized by chaotic velocity patterns~\cite{carenza19turb}. The following Sections will be devoted to present the different behaviours of the sheared system at varying both the intensity of active and external forcing.

\subsection{Linear velocity profiles and lamellar phase.}
The scenario just described is strongly influenced by an external shear flow. Due to the tendency of lamellae to align with the flow, an applied shear, even small, is found to counter activity-induced bending, thus extending the range of stability of  lamellar order   towards larger $Er_{act}$ ($Er_{act} \lesssim 0.18$) with respect to its unsheared counterpart. Under this threshold and for a vast range of shear rates, the system sets into a lamellar phase, as shown in panels (a) and (d) of Fig.~\ref{fig:S1}. The region with these properties is red in Fig.~\ref{fig:phase_diagram}, where flow regimes found by scanning the $Er-Er_{act}$ plane are summarized. At small shear, relaxation dynamics  leads to the formation of long-lived dislocations  in the lamellar pattern, as the one highlighted by the black box in panel (a) of Fig.~\ref{fig:S1}. Panel (b) of the same Figure shows the detail of the velocity field in the neighborhood of the dislocation. If $\dot{\gamma}$ is weak enough, defects are capable to consistently alterate the velocity pattern, since dislocations develop flows trasversal to the direction of lamellar-alignment, thus leading to permanent shear bandings in the velocity profile (as shown by the red line in the inset of Fig.~\ref{fig:S1}c). At greater values of shear rate, the superimposed flow is strong enough to eliminate dislocations (see for example panel (d) in Fig.~\ref{fig:S1} and Movie~1), eventually leading to the formation of disruptions that are much less effective on the flow than dislocations, as confirmed by the linear behaviour of the corresponding velocity profile. In this regime lamellae are globally aligned to the flow, while the polarization field, homeotropically anchored to the interfaces (panel~(e) of Fig.~\ref{fig:S1}), is pointing towards the passive phase (blue regions in zoom of panel~(e)). Panel~(f) compares the time evolution of shear stress for the two cases considered. Dynamics at high shear leads to a smoother and faster relaxation towards lower values of $\sigma_{xy}$. When the imposed shear is weaker, oscillations or  jumps, as the one marked by an arrow at $t=6\times 10^6$ in panel (f) of Fig.~\ref{fig:S1}, are due to the annihilation of two dislocations.

\subsection{Unidirectional motion.}
The behaviour becomes  more complex when activity is increased. The combination of activity and shear has dramatic consequences.
The system undergoes a morphological transition from the lamellar phase towards an emulsion of active material in a passive background, a behaviour also found by Bonelli \emph{et al.}~\cite{bonelli2019} at lower active dopings. Fig.~\ref{fig:S3}  shows two cases  at $Er=0.031$, $Er_{act} = 0.29$ (top row) and  $Er=0.062$, $Er_{act} = 0.29$ (bottom row) characterized by the formation of a thick  layer of material close to \textit{one} boundary  (see dynamics in Movie~2), and small features on   Brazovskii lengthscale $\lambda=2\pi/\kappa$ coexisting  with larger aggregates of active material elsewhere.
Such symmetry breaking is mediated and sustained by the formation of these large active domains where bending polarization instabilities, typical of extensile systems, act as a source of vorticity (see panel~(b)). Big active domains are mostly advected by the intense flow close to the walls, as shown by purple velocity streamlines, differently from what happens in shear-free systems where polarization bending results into the rotational motion of the bigger active droplets. Fig.~\ref{fig:S3}c shows the related $x$-averaged velocity profile $\langle \mathrm{v_x}\rangle$ (grey curve): Instead of the linear behaviour  of  Fig.~\ref{fig:S1}c, one observes banded flows with  the higher gradient in correspondence of  the wall with the active layer.
Similar cases occur  at different $Er, Er_{act}$ with deposition of active material randomly on the top or bottom wall. Streamlines of $\textbf{v}$ (in panels (b) and (e) of Fig.~\ref{fig:S3}) show that the inversion of the fluid velocity takes place in correspondance of the interface of the active layer. The top-bottom symmetry breaking leads to a net flux of matter in the flow direction and has been named \textit{unidirectional motion}. Cases exhibiting such property have been plotted in grey in Fig.~\ref{fig:phase_diagram}. Moreover, as shear is increased, the position of flow inversion migrate towards the bulk of the system (see the dark grey profile at the larger $Er$ in panel (c) and the corresponding $\mathrm{v_x}$ contour plot in panel (f)). Unidirectional flow (grey region in Fig.~\ref{fig:phase_diagram}) may occur with  almost everywhere vanishing gradient of $\langle \mathrm{v_x}\rangle$, and in this case it will  be called  \textit{superfluidic}~\cite{cates2008,nota_guo} (see  light grey profile  in Fig.~\ref{fig:S3}c).

Which are the mechanisms for  the observed velocity profiles? And how to explain the flow symmetry breaking transition? Due to complexity of the system we can only partially answer to these questions. The velocity behaviour is strictly related to that of  polarization close to the walls. Thick layers  as the one  in Fig.~\ref{fig:S3}a,d and  in Fig.~\ref{fig:pol_at_walls} are characterized by the bending of polarization due to competition between strong parallel anchoring to the walls  and perpendicular orientation to domain interfaces, leading to a negative active shear stress contribution. This is clearly shown in Fig.~\ref{fig:pol_at_walls},  where white/black regions correspond to positive/negative values and correspond to active domains, while beige ones are associated to the isotropic background and correspond to almost null values. Moreover, topological defects in the active layer are strongly inhibited by elastic energy, as suggested by the uniform polarization pattern in the black bottom layer.

Within the active layer $\sigma_{xy}^{act} \sim \frac{\zeta}{2}\phi_0 (P_{eq})^2 \sin{2\theta}$, where $\theta$ denotes the local orientation of polarization with respect to the imposed velocity ($ 0\le \theta \le \pi$), thus generating an active force density in the flow direction ($f^{act}_\parallel= \partial_\perp \sigma_{xy}^{\textrm{act}}$, where $\partial_\perp$ denotes derivative in the direction normal to the walls). This can either reinforce the imposed flow if the polarization field is oriented as $v_w$ (since $\partial_\perp \sin 2\theta > 0$), or lead to a reduction of the fluid velocity if opposite (since $\partial_\perp \sin 2\theta < 0$). However, between the two possible orientations, the one reinforcing the flow does not appear in the cases discussed so far. In order to clarify this point we define the average polarization $P_w$ on each wall -- it can be calculated as $P_w\sim \langle P_\parallel \rvert_\textrm{walls} \rangle $, where here $\langle \cdot \rangle $ stands for the average over few layers close to  the wall sites. For the unidirectional motion case, $P_w$ is null at the wall where the thick active layer is absent, so that we denote such polarization state as $\mathbf{0}$. If the polarization is opposite to $v_{w}$, as in  the bottom/top of panel (b)/(e) of Fig.~\ref{fig:S3}, we will indicate it  with $ \textbf{-}$. Following the notation just introduced, we will refer to the global states shown in Fig.~\ref{fig:S3} with $\mathbf{0} \textbf{-}$, independently of the top-bottom asymmetry.
\noindent

\subsection{Symmetric shear thinning profiles.}

By further increasing both activity and shear rate, phase demixing is more pronounced  (Fig.~\ref{fig:S4}a) with the formation of an emulsions of amorphous active domains in a passive matrix.
Active layers form on \textit{both} walls so that symmetry is  restored also at level of the velocity profiles $\langle v_x \rangle $, with gradients in the bulk of the system lower than the imposed one, as  shown  in panel~c of Fig.~\ref{fig:S4} (thick lines).
Symmetric cases exhibiting such phenomenology have been plotted in green in Fig.~\ref{fig:phase_diagram}.
Under these conditions, it may happen that the velocity gradient in the bulk of the system is either  everywhere vanishing or, eventually, opposite to the one externally imposed (\textit{negative viscosity}), despite such states are found to be unstable in the long term  (see Sec.~\ref{sec:intermittent}).
To explain the flow properties presented, we analyse the active shear stress profiles $\langle \sigma_{xy}^{act} \rangle$ averaged in the flow direction (see thin lines in panel~(c)). This confirms that the active stress is considerably different from zero only in the layer close to both walls, where it assumes negative values --thus leading to the sharp decrease of the intensity of the flow in the same region --  while it is approximately null in the bulk. A contour plot of the active stress is also shown in panel~b of Fig.~\ref{fig:S4}, clearly showing that the $P_w$ polarization state at boundaries is in a $\textbf{--}$ configuration.

\begin{figure}[t]
\centering
{\includegraphics[width=.46\textwidth]{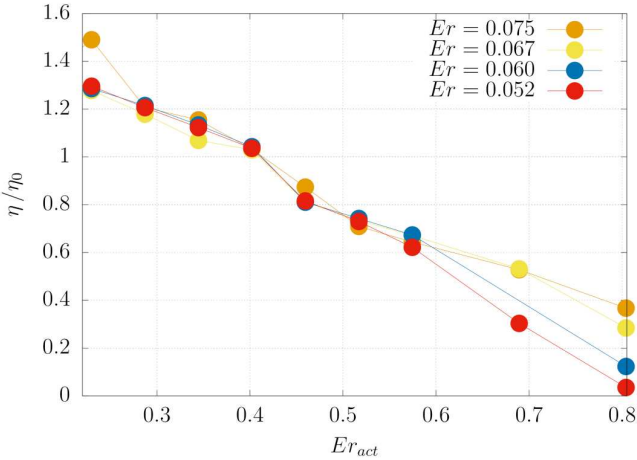}}
\caption{\textit{Shear thinning in extensile mixtures}. Ratio between apparent viscosity $\eta$ (measured as $\langle \sigma^{tot}_{xy} \rangle /\dot{\gamma}$) and shear viscosity $\eta_0$, while varying activity for some values of the Ericksen number.
}
\label{fig:costitutive}
\end{figure}

Such behaviour is also accompanied by shear thinning, typical of extensile fluids as $Er_{act}$ is increased. This is  analyzed in Fig.~\ref{fig:costitutive} where the ratio between the apparent viscosity $\eta=\langle \sigma^{tot}_{xy} \rangle / \dot{\gamma}$  (where $\langle \sigma^{tot}_{xy} \rangle$ denotes the time average of the total stress tensor),  and the shear viscosity $\eta_0$ has been plotted versus $Er_{act}$. We varied  $Er$ in the range   $Er\gtrsim 0.05 $, where viscosity states are found to be stable for any value of $Er_{act}$. Viscosity mainly depends on the intensity of the active doping, while no substantial dependence is found on the shear rate if $Er_{act} \lesssim 0.6$. This suggests that  activity, inducing shear thinning, is a parameter capable of controlling the rheological property of extensile suspensions.

\begin{figure*}[t]
\centering
{\includegraphics[width=.75\textheight]{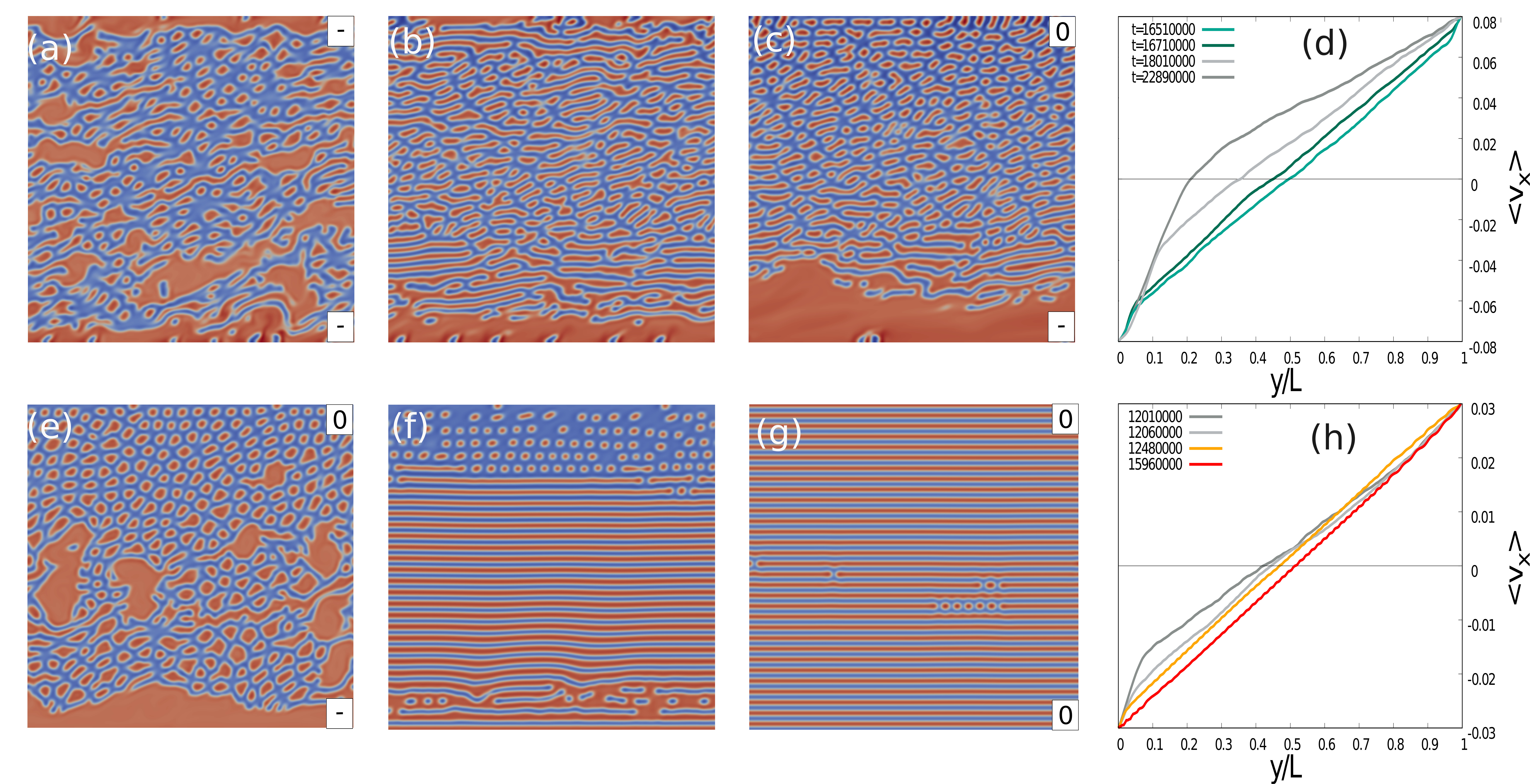}}
\caption{\textit{Activity quench}. 
Panels (a)-(c) show the evolution of the system after that $Er_{act}$ has been quenched from $0.57$ to $0.28$, thus moving from the green region of Fig.~\ref{fig:phase_diagram} to the grey one, at $Er=0.06$. 
The colors of the profiles have been chosen in accordance to the color of the corresponding region in the phase diagram of Fig.~\ref{fig:phase_diagram}.
Labels ($\textbf{0}$ or $\textbf{-}$) at the top and at the bottom of initial and final panel denote the state of the polarization at the boundary. The evolution of velocity profiles is also shown in panel~(d).
Panels (e)-(g) show the same for the quench from the grey region ($Er_{act}=0.22$) towards the red region ($Er_{act}=0.057$) of Fig.~\ref{fig:phase_diagram} at $Er=0.01$, with the evolution of velocity profiles in panel~(h).
Contour plots correspond to the first, third and last  velocity profiles in panels (d) and (h)}.
\label{fig:quench_1}
\end{figure*}

\subsection{Activity quench.}
\begin{figure*}[t]
\centering
{\includegraphics[width=0.97\textwidth]{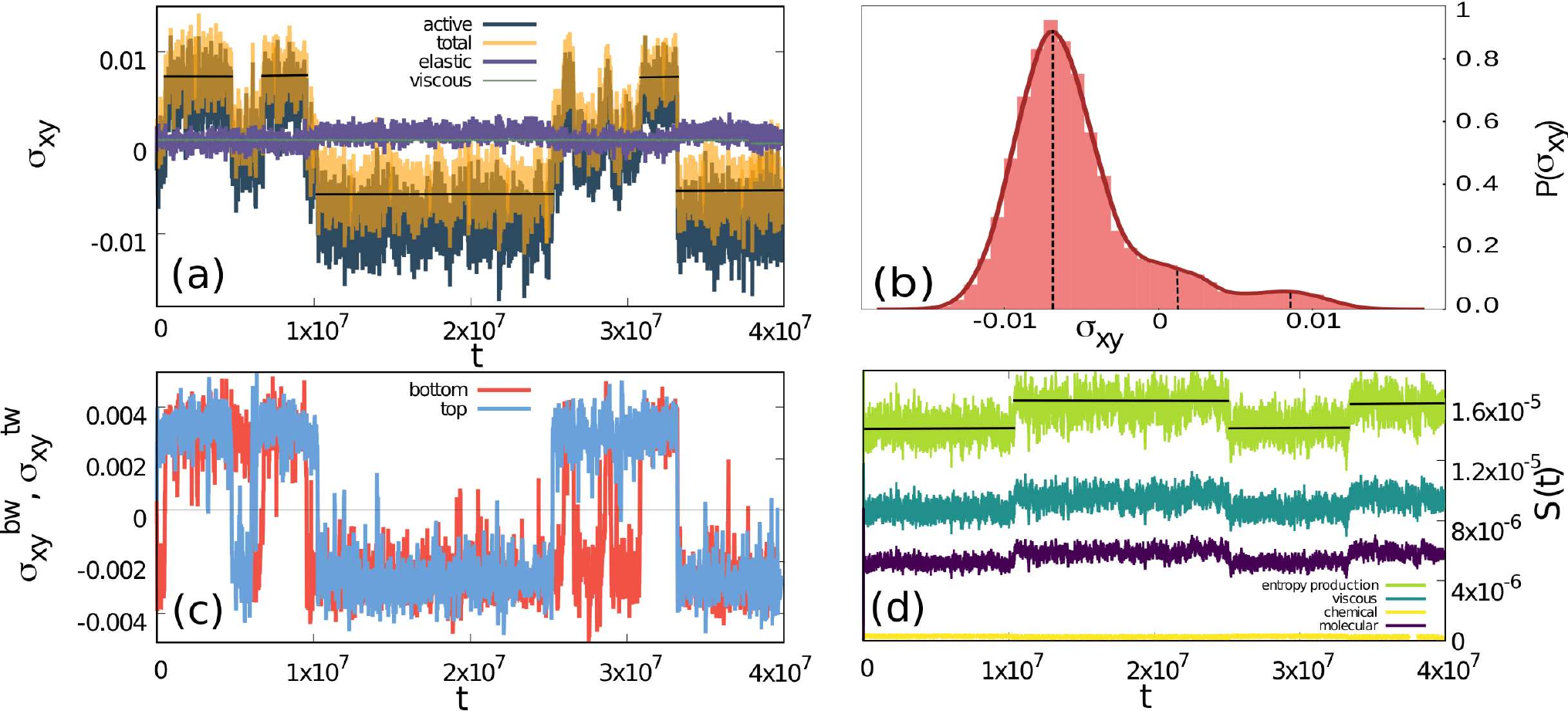}}
\caption{ \textit{Entropy production in multistable states.} (a) Time evolution of stress contributions at $Er_{act}=1.14$,$Er=0.00075$. The total stress is plotted in transparent yellow, while the active one is the dark blue curve underneath. Panel (b) shows the \textit{pdf} of the total stress (data from 40 independent runs, fitted by 3 normal distributions peaked at $\sigma_{xy}=-0.0068,0.0009,0.0086$). (c) Total stress close to bottom and top walls, respectively computed as $\sigma^{tw}_{xy} = \int dx \int^{Ly}_{Ly-l}  dy \ \sigma_{xy}$ and  $\sigma^{bw}_{xy} = \int dx \int^{l}_{0}   dy \  \sigma_{xy}, $ where $l$ is the width of the layer. Here $l=15$. Nevertheless, as long as $l<Ly/2$, results remain unaltered. (d) Entropy production. Negative viscosity states, corresponding to $\textbf{-}\textbf{-}$ configurations of polarization at the boundaries, live longer than others. Entropy production $s(t)$ assumes greater values in correspondence of these regions. Each contribution to $s(t)$ has been measured by integrating on the computational domain the terms on the right-hand side of Eq.~\eqref{eqn:entropy_production}.
}\label{fig:S5}
\end{figure*}

We further analyzed the nature of the transition between the symmetric configurations at higher activity (green region in Fig.~\ref{fig:phase_diagram}) and states with unidirectional flow at weaker activity, starting from a stationary states at $Er_{act}=0.57$ and quenching the activity to $Er_{act}=0.28$ at fixed $Er=0.06$. Panels (a)-(c) in Fig.~\ref{fig:quench_1} show the quenching dynamics: starting from the symmetric configuration of panel~(a) at $Er_{act}=0.57$, amorphous active domains progressively stretch in the flow direction and cluster on the bottom boundary (see panel~(b)). This accompanies the \textit{melting}   of the active layer close to  the upper wall, finally generating the asymmetric configuration of panel~(c) characterized by unidirectional flow. The evolution of velocity profiles is shown in panel~(d). Panels (e)-(g) show the results of a similar experiment: this time we quenched the active parameter so to move from the grey region ($Er_{act} = 0.22 $) with unidirectional motion, to the red  one ($Er_{act} = 0.057 $) with linear profiles, in Fig.~\ref{fig:phase_diagram}, while keeping shear rate fixed at $Er=0.01$. After the quench droplets are no more stable, since the active doping is not strong enough to maintain the bending instability. A lamellar phase progressively grows from the bottom of the system, where most of the active material was initially found, towards the upper wall. The final configuration is characterized by a symmetric, defect-free lamellar configuration, with linear velocity profile, whose evolution can be appreciated by looking at panel~(h).

\section{Intermittent flow}
\label{sec:intermittent}
Surprisingly, yet another behaviour appear in a vast region of the parameter plane at small shear, $ Er \lesssim 0.04$, and sufficiently large activity, $Er_{act} \gtrsim 0.4$,~(blue in Fig.~\ref{fig:phase_diagram}). Under this condition, the system is symmetric and exhibits an intermittent flow regime.
This is to be related, once again, to the polarization state at the boundaries. Indeed, this time we find (unstable) configurations where $P_w$ is oriented in the same direction of the imposed velocity, so that we will denote such state as $\textbf{+}$. The dynamics of the system is characteried by jumps between $\textbf{+}\textbf{+}$, $\textbf{-} \textbf{+}$ and $\textbf{-} \textbf{-}$ states.

Such intermittent behaviour is reflected in the evolution  of the area averaged stress, as shown in Fig.~\ref{fig:S5}a. Elastic contributions  are on average constant, while active stress fluctuates around positive, negative or vanishing values. These are found to be largely determined by the portion of the system closer to boundaries (see Fig.~\ref{fig:S5}c). For each wall the sign of active stress coincides with the one of $P_w$, so that positive and negative total stress correspond respectively to $\textbf{+} \textbf{+}$ and  $\textbf{-} \textbf{-}$  $P_w$ states, while total zero active stress comes from opposite $P_w$ contributions ($\textbf{-+}$ states). Viscosity jumps~(Fig.~\ref{fig:S5} a) correspond to the inversion of the polarization on one of the two walls during evolution. The inversion of polarization on the boundaries, generally prevented by elastic effects and  strong anchoring, can  occur as a result of \emph{catastrophic} events, such as collision of big  domains with the active layer (see Appendix~B, and  Movie~3,4). $\textbf{-} \textbf{-}$ states typically live  longer than others, as can be appreciated looking at the  \textrm{pdf}  of the total shear stress reported in Fig.~\ref{fig:S5}b.  The statistics of viscosity states has been constructed considering 40 different runs for the same couple of parameters {$Er_{act}=1.14$, $Er=0.0075$}. The  typical length of each run is of about $10^7$ lattice Boltzmann iterations. The data for the shear stress have been sampled under stationary conditions and then  fitted with the sum of three normal distributions, centred at values marked with dotted lines in panel~(b). These mean values are consistent with the average of the stress in the different states of Fig.~\ref{fig:S5}a, and their probability of occurrence will correspond to the amount of time that the system spends in each of them.

To go deeper in the characterization of this behaviour we measured entropy production during the system time evolution. Generally, in non-equilibrium systems, the entropy density $\Sigma(\vec{r},t)$ obeys the continuity equation \begin{equation}
\label{eqn:entropy_continuity_equation}
\partial_t \Sigma + \nabla \cdot (\Sigma \vec{v}) = s,
\end{equation}
where $s$ is  the rate of entropy  production per unit volume, subject to the condition $s \geqslant 0$. This can be written in terms of generalized fluxes $J_i$ and forces $X_i$, as~\cite{degroot1962}
\begin{equation}
T s = J_i X_i,
\label{eqn:entropy_production_general}
\end{equation}
with $T$ the temperature -- fixed at in our simulations ($T=0.5$) since we are neglecting heat transfer. Thermodynamic forces are chosen as follows:
\begin{align}
\tilde{X}_v &= (\nabla \vec{v})^S \equiv \tilde{D}, \\
\vec{X}_P &=  - \dfrac{\delta F}{\delta \vec{P}} \equiv \vec{h}, \\
\vec{X}_\phi &= \nabla  \dfrac{\delta F}{\delta \phi} \equiv \nabla \mu,
\label{eqn:forces}
\end{align}
where $F$ is given by Eq.~\eqref{eqn:fe}, $\tilde{D}$ is the strain rate tensor. Moreover, in our model the following linear phenomenological relations between forces and fluxes hold~\cite{kruse2004}:
\begin{align}
\tilde{J}_v &= \tilde{\sigma}^{tot}, \\
\vec{J}_P &= \partial_t \vec{P} + (\vec{v} \cdot \nabla) \vec{P} + \tilde{\Omega} \cdot \vec{P} = \dfrac{1}{\Gamma} \vec{h} + \xi \tilde{D} \cdot \vec{P}, \\
\vec{J}_\phi &= -M \nabla \mu,
\label{eqn:fluxes}
\end{align}
where we denoted with $\tilde{\Omega}$ the vorticity tensor, that is the antisymmetric part of the gradient velocity tensor. By substituting these relations into Eq.~\eqref{eqn:entropy_production_general}
we find that
\begin{equation}
\label{eqn:entropy_production}
T s = 2 \eta \tilde{D} : \tilde{D} + \dfrac{1}{\Gamma} \vec{h} \cdot \vec{h} + M (\nabla \phi)^2,
\end{equation}
where we retained only those terms even under time reversal symmetry. We here identify three contributions: the first one is the entropy production due to viscous effects, $2\eta  \tilde{D} : \tilde{D}$,
while the second  ($\frac{1}{\Gamma} \vec{h} \cdot \vec{h}$) and the third ($M (\nabla \phi)^2$) ones are respectively the molecular and the chemical terms, accounting for the entropy produced during the relaxation dynamics of $\vec{P}$ and $\phi$.
This expression is quadratic in the thermodynamic forces and
satisfies all required conditions, among which
invariance under Galilean transformations as well as the second principle of thermodynamics.
In the stationary regime, entropy production must be equal to the energy injected into the system, due to the work of the walls and active pumping.
We emphasize that activity $\zeta$ is a reactive parameter and therefore does not appear in the entropy production formula~\cite{kruse2005}.
However activity influences the dynamics, acting as a velocity source through the active stress, thus contributing indirectly to dissipation.

Panels (a) and (d) compare the  behaviour of the  stress and entropy production. The evolution of the stress is characterized by two regions of stability of negative viscosity -- whose lifetime is $\mathcal{O}(10^7)$ LB iterations -- that respectively occur  from $t \approx 10^7$ to $2.5\times 10^{7}$, and from $t \approx 3.2\times 10^{7}$ to $4\times 10^{7}$. These states have also the highest probability, as can be appreciated looking at the pdf of the total stress in panel (b) of Fig.~\ref{fig:S5}. 

In Panel~(d) the different entropy production contributions are shown. We first notice that the contribution due to diffusion/chemical (yellow line) is almost null. In addition, the viscous dissipation~(blue line) is always greater than the contribution due to the molecular field~(violet line). This suggests that the hydrodynamics of the system --  driven by active injection and external forcing -- is mainly countered by viscous dissipation phenomena. Moreover, the total entropy production oscillates around two different values and jumps during time evolution, with the highest value corresponding to the negative viscosity states.
This behaviour, with prevalence of $\textbf{-} \textbf{-}$ polarization states is typical of all the intermittent cases (blue region in Fig.~\ref{fig:phase_diagram}), and is compatible with a  maximum entropy production principle~(\emph{MaxEPP}).
Various variational principles, related to entropy production rates, have been put forward to quantitatively select the most probable state in multi-stable systems. While 
much efforts have been spent in the search of a general principle and recent progress has been made~\cite{dewar2013beyond} on a theoretical derivation of such a principle, questions about how this  should be interpreted and applied have not been answered, especially for systems evolving far from thermodynamic equilibrium.  
In particular, a \emph{MaxEPP} has been implemented as a selection criteria to study systems characterized by multiple non-equilibrium stationary states~\cite{dewar2013beyond} and very recently the Schl\"{o}gl model~\cite{Schlgl1972} -- that is a simple, analytically solvable, one-dimensional bistable chemical model -- has been used to demonstrate that the steady state with the highest entropy production is favoured~\cite{endres2017}. 
%Our simulations suggest that in the system here considered entropy production
In the system here considered, we found that entropy production $s(t)$ is higher  for the most likely $--$ states, suggesting that \emph{MaxEPP} may act as a thermodynamic principle in selecting non-equilibrium states.

\section{Overview, phase diagram and conclusions}
\begin{figure}[t!]
\centering
{\includegraphics[width=.47\textwidth]{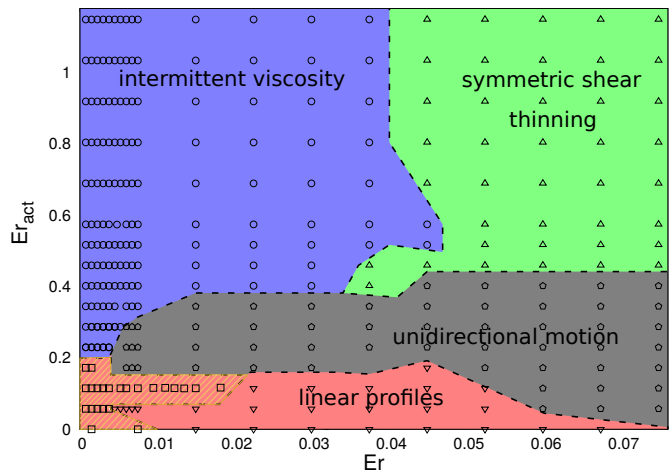}}
\caption{Flow regimes in the $Er$-$Er_{act}$ plane,
as described in the text.
Hollow marks denote  simulation points.}
\label{fig:phase_diagram}
\end{figure}

The various behaviours found in the active polar emulsion under shear have been summarized in Fig.~\ref{fig:phase_diagram}, at varying $Er$ and $Er_{act}$.
At small $Er_{act}$, the system arranges in a lamellar configuration. In this range, at small shear rates, few persistent defects (dislocations) give rise to slight deformations in velocity profiles (dashed-red region in Fig.~\ref{fig:phase_diagram}). Increasing $Er$, dislocations are washed out by the flow and the system enters the region of linear velocity profiles (red region in Fig.~\ref{fig:phase_diagram}). As activity is increased, morphology is  characterized by a transition towards asymmetric configurations, with the formation of a thick  layer of material close to \textit{one} boundary, thus generating a non-vanishing flux of matter  -- a behaviour addressed as \emph{unidirectional flow}. Such regime is stable for a broad range of $Er$ at intermediate active dopings (grey region in Fig.~\ref{fig:phase_diagram}).
By further increasing activity, at high shear rate, phase demixing is more pronounced  and this has important consequences on the flow. Active layers form on \textit{both} walls and symmetry is restored. Under these conditions, it may happen that the velocity gradient in the bulk of the system is either  everywhere vanishing or, eventually, opposite to the one externally imposed (\textit{negative viscosity}), despite such states are found to be unstable in the long term.  The region with stable symmetric profiles is  green in Fig.~\ref{fig:phase_diagram}.
Cases in the blue region are instead characterized by the jumping dynamics described in the previous section. 

We also remark that the results presented in this paper strictly hold for bidimensional geometries -- as often happens in experimental realizations of active systems, where bacteria and cytoskeletal suspensions are usually confined at a water-oil interface \cite{wensink2012,doostmohammadi2018}.
In full $3d$ environments -- where both vortex stratching of the flow field and twisting of the polarization field are allowed -- the proliferation of degrees of freedom may strongly affect the behaviour of the system that is indeed different even in absence of any internal and/or external forcing~\cite{Henrich2012,carenza2019b}.

In conclusions, we showed how the competition between externally imposed shear and local
energy injection results in a wealth of different rheological behaviours, that can be explained in terms of specific dynamical mechanisms. As an example, jumps between velocity profiles with positive and negative gradients are due to collisions between large active droplets or domains and active layers coagulated on the moving walls.
The  generalized active gel model proposed in this work has
%theory that we implemented to perform the study 
allowed us to perform a fully $2d$ analysis, by keeping under control the time evolution of the important variables, such as the local concentration and the  orientation of the active constituents. Thus we confirmed, by varying both external and internal forcing, the existence of superfluidic and negative viscosity states found experimentally in bacterial suspensions~\cite{lopez2015,guo2018} whose first numerical confirmation by means of quasi-$1d$ simulations was furnished in~\cite{loisy2018}. Moreover, we also found that a
maximum entropy production principle holds in selecting the most probable state in the intermittent viscosity regime.
Since most of the observed behaviours mainly arise due to the elastic properties of the  order parameter,
they are expected to stay valid also in nematic systems. This because the dynamics of polar systems differs from their nematic counterpart mainly for the allowed topological defects.
We hope that this study can stimulate the design of new active materials and devices with pioneering applications.

\section*{Conflict of Interest}
There are no conflicts of interest to declare

\section*{Acknowledgments}
Simulations were  performed at Bari ReCaS e-Infrastructure funded by MIUR through the
program PON Research and Competitiveness 2007-2013 Call 254 Action I. A.T. acknowledges funding from the European Research Council under the European Union's Horizon 2020 Framework Programme (No. FP/2014-2020) 
ERC Grant Agreement No.739964 (COPMAT). We thank Davide  Marenduzzo and Ilario Favuzzi for the  useful discussions.

%%%REFERENCES%%%
\bibliography{refs} %You need to replace "rsc" on this line with the name of your .bib file
\bibliographystyle{unsrt}

\appendix

\section{Adimensional numbers}
\label{sec:adimensional_numbers}
In this Appendix we will furnish a derivation for the compression modulus $B$ that we made use of to define the adimensional Ericksen number, $\textrm{Er}$, and the \emph{active} Ericksen number $\textrm{Er}_{act}$.

In the following we consider the elastic coefficients for a binary mixture in the lamellar phase,  in which one of the component is an isotropic fluid and the other is a polar liquid crystal.
One of them, the compression modulus, is used to define the adimensional Ericksen number $Er$
and its active counterpart $Er_{act}$, in terms of the model parameters. The analytical treatment generalizes that given in~\cite{jaju2016} for a simple lamellar fluid.

It is first  convenient to rewrite the Landau-Brazovskii free-energy functional of Eq.~\eqref{eqn:fe} in a more symmetric form, in terms of the  field $\psi = \phi-\phi_{cr}$, as
\begin{eqnarray}
\label{eqn:fe2}
F[\psi,\mathbf{P}]=
\int d\mathbf{r}\,\left[ \frac{\tilde{a}}{2} \psi^2 + \frac{b}{2} \psi^4 +  \frac{k_\phi}{2}\left|\nabla \psi\right|^{2}+\frac{c}{2}(\nabla^2\psi)^2 \right. \nonumber\\ \left.
-\frac{\alpha}{2} \psi \left|\mathbf{P}\right|^2+ \frac{\alpha}{4}\left|\mathbf{P}\right|^{4}+\frac{k_P}{2}(\nabla\mathbf{P})^{2}
+\beta\mathbf{P}\cdot\nabla\psi \right] \ \ .
\end{eqnarray}
Eq.~\eqref{eqn:fe} of the main text can be obtained with $\tilde{a}= - a/\phi_{cr}^2$, $b=a/\phi_{cr}^4$ and $\phi_0=2 \phi_{cr}$. At equilibrium, the chemical potential $\mu$ and the molecular field $\bf{h}$ must vanish:
\begin{align}
\mu \equiv \frac{\delta F}{\delta\psi}&= \tilde{a} \psi + b \psi^3 -k_\phi \nabla^2 \psi + c \nabla^4 \psi -\alpha{\bf P} - \beta \nabla \cdot {\bf P} =0, \label{eqn:phi_equilibrium} \\
\bf{h} \equiv \frac{\delta F}{\delta{\bf P}}&= -\alpha \psi {\bf P} + \alpha {\bf P}^2 {\bf P} -k_P \nabla^2 {\bf P} + \beta \nabla \psi=0. \label{eqn:P_equilibrium}
\end{align}
We then take the  single mode approximation~\cite{jaju2016}, exact if considering  only gradient terms in the above expressions,
\begin{equation}\label{eq_psi}
\psi= \tilde{\psi} \sin ( \kappa y),
\end{equation}
where the amplitude $\tilde{\psi}$ and the wavenumber $\kappa$ of the modulation have to be computed. Our simulations  confirm that, as long as bulk parameters are small if compared to the elastic ones, the concentration field is modulated in a sinusoidal fashion.
By substituting Eq.~\eqref{eq_psi} into Eq.~\eqref{eqn:P_equilibrium}, and neglecting non linear contributions, we find that $P_y$ must satisfy,
\begin{equation}
\partial_y^2 P_y = - \frac{\beta \kappa}{k_P}  \tilde{\psi} \sin ( \kappa y),
\end{equation}
whose periodic solutions, with lamellar width $\lambda=2\pi/\kappa$, are given by
\begin{equation}
P_y= C + \dfrac{\tilde{P}}{\kappa} \cos ( \kappa y),
\end{equation}
where $C$ is a constant  and
\begin{equation}
\tilde{P}= \beta \tilde{\psi}/k_P.
\label{eqn:polarisation_amplitude}
\end{equation}
In order to find the coefficients $C$, $\tilde{\psi}$ and the wavenumber $\kappa$, we substitute profiles of $\psi$ and $P_y$ in Eq.~\eqref{eqn:fe2}, integrate over the lamellar wavelength $\lambda$ and minimize with respect to $\kappa$, that is found to be $\sqrt{|k_\phi|/2c}$, the same as for the polarization-free case. Thus we rewrite the free-energy density as
\begin{eqnarray}
  f = \dfrac{1}{4} \left[ ( \tilde{a}-a_\phi -a_P)\tilde{\psi}^2 + \dfrac{3}{8}\left(  b   + \dfrac{4 \alpha c^2 \beta^4}{ k^2_\phi k^4_P} \right) \tilde{\psi}^4  \right. \nonumber \\ \left.
  + \alpha C^2 \left( C^2 - \dfrac{6 c \beta^2 \tilde{\psi}^2}{k_\phi k_P^2} \right) \right] ,
\end{eqnarray}
where we defined $a_\phi=k^2_\phi/4c$ and $a_P= \beta^2/k_P$.
Minimization of $f$ with respect to $C$ gives $C=0$.
Then, by further minimizing $f$ with respect to $\tilde{\psi}$, we find
\begin{equation}
\tilde{\psi} = \pm 2|k_\phi| k_P^2 \sqrt{\dfrac{a_\phi + a_P - \tilde{a}}{3(k^2_\phi k_P^4 b + 4 \alpha c^2 \beta^4)}}.
\end{equation}
This result shows that lamellar ordering occurs for $\tilde{a}< a_{cr} = a_\phi + a_P $. Thus, polarization enlarges the range of stability of the lamellar phase with respect to the Brazovskii theory, where lamellar ordering occurs if $\tilde{a}<a_\phi$.

We now introduce the elastic coefficients related to the free-energy cost of deviations
from the harmonically modulated profile.
We perturb equilibrium profiles by introducing a layer perturbation field $\vartheta(x,y)$, in terms of which
the perturbated profiles become
\begin{eqnarray}
 \label{psi_f}\psi(x,y) &=& \tilde{\psi} \sin \left[ \kappa(y-\vartheta(x,y)) \right],\\
 P(x,y) &=& \dfrac{\tilde{P}}{\kappa} \cos \left[ \kappa(y-\vartheta(x,y)) \right].\label{P_f}
\end{eqnarray}
The field $\vartheta$ is chosen so that its amplitude is much smaller than the lamellar width $\lambda$, but its typical variation lengthscale is much wider.

\begin{figure*}[t]
\centering
{\includegraphics[width=1.0\textwidth]{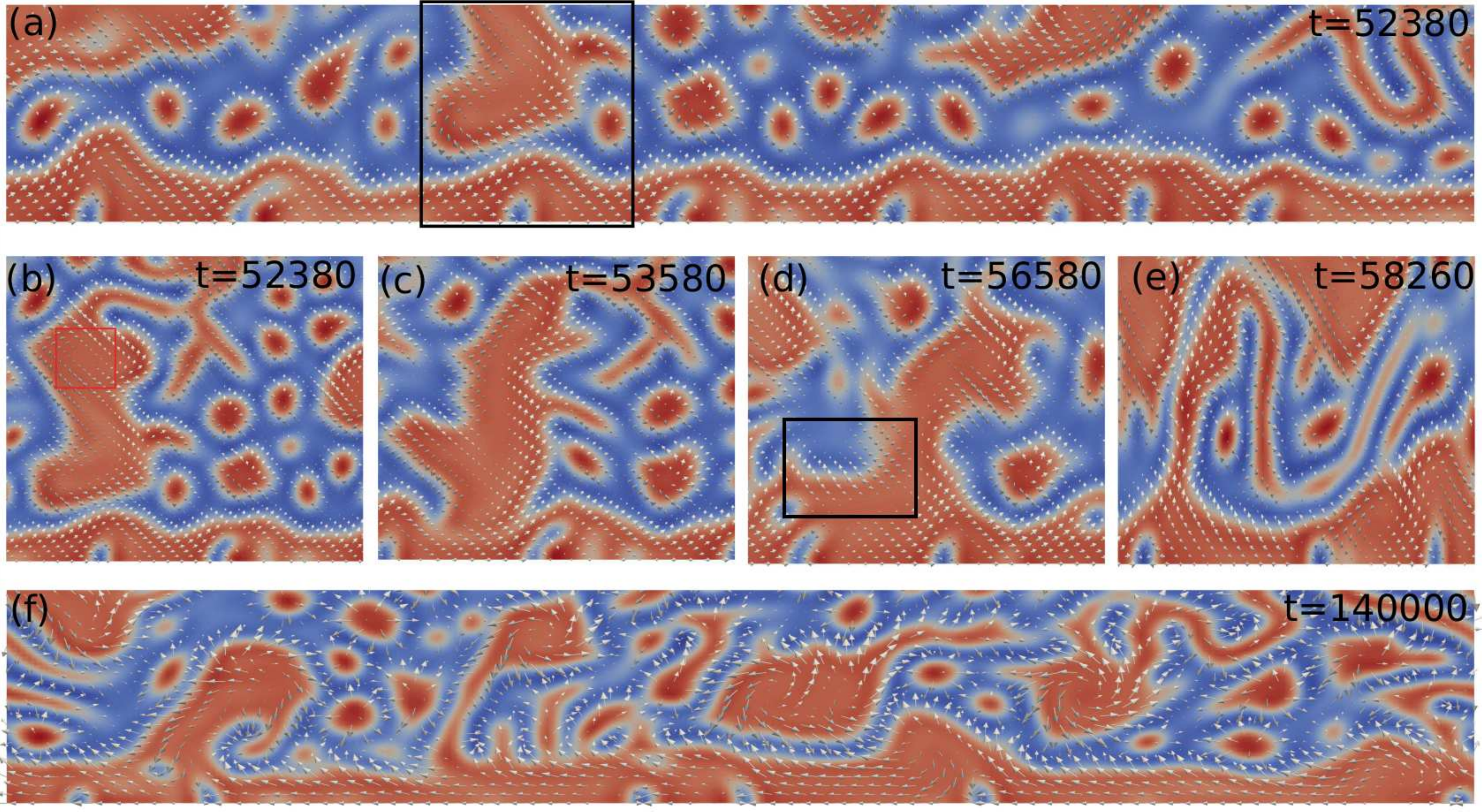}}
\caption{\textit{Polarization flip.} Contour plots of concentration field $\phi$ for $\zeta=0.005$ ($Er_{act}=0.57$) and $\dot \gamma=3.12\times 10^{-5}$($Er=0.003$). Panels (a) and (f) show the initial and final state, respectively, of the polarization at the bottom wall. Panels (b-e) show snapshots of the collision of a droplet with the bottom wall, which causes a local rearrangement of the polarization state. The red contour in panel~(b) encloses a $+1$ defect in the polarization field pattern.}
\label{fig:S2}
\end{figure*}

Because of the slowly-varying behaviour of $\vartheta$, we flush out high-frequency modes to obtain a coarse-grained description of the model solely in terms of the layer displacement.
This implies~\cite{jaju2016} that the following normalization condition holds:
\begin{equation}\label{norm}
\left[ \dfrac{1}{L_x L_y} \int d \mathbf{r} \ \sin^2 \left[ \kappa(y-\vartheta(x,y) \right] \right]_{cg} = 1,
\end{equation}
where $L_x$ and $L_y$ are the linear dimensions of the system and the boundary conditions are assumed to be periodic in both directions.
By substituting Eqs.~\eqref{psi_f}-\eqref{P_f} and their derivatives in Eq.~(\ref{eqn:fe2}) and by retaining only the elastic contributions, namely the gradient terms, we find the following coarse-grained free-energy functional:
\begin{multline}
 F_{cg} \left[ \vartheta \right] = \int d \mathbf{r} \ \dfrac{\tilde{\psi}^2}{2} \left[ \left( k_\phi \kappa^2 + 2c \kappa^4 + \dfrac{\beta^2}{k_P} \right)(\partial_x \vartheta )^2 \right. \\
    %\right. \qquad \qquad \\
    %\qquad \left.
    \left.+\left( k_\phi \kappa^2 + 6c \kappa^4 + \dfrac{\beta^2}{k_P} \right)  (\partial_y \vartheta )^2 + c \kappa^2 (\nabla^2 \vartheta)^2 \right] \\
    \equiv \int d \mathbf{r} \ \left[  \dfrac{\Sigma}{2} (\partial_x \vartheta )^2 + \dfrac{B}{2} (\partial_y \vartheta )^2 + \dfrac{\Upsilon}{2} (\nabla^2 \vartheta)^2 \right]
\label{eqn:free_energy_coarse_grain}
\end{multline}
where we have used Eq.~\eqref{norm}, and Eq.~\eqref{eqn:polarisation_amplitude} to get rid of $\tilde{P}$.
In this expression we identify  three  contributions,
%Any deviation from such configuration can be seen as the
combination of the three following effects: \textit{(i)} stretching/shrinking of the lamellar surface in the layer direction ($x$ in this Section) ,  \textit{(ii)} compression/expansion of the lamellar layers in the gradient direction ($y$) and \textit{(iii)} bending of the layers.

The coefficient of the derivative along the gradient direction is half of the compression modulus $B$ and it gives the energy penalty per unit surface due to a change in the layer width. Its explicit expression is then given by
\begin{equation}
B=\left(\dfrac{\beta^2}{k_P} + \dfrac{k_\phi^2}{c} \right) \tilde{\psi}^2.
\label{eqn:compression_modulus}
\end{equation}
Analogously we can define the surface tension $\Sigma$ as the energy penalty per unit surface due to the stretching of the layer as half of the coefficient in the layer direction
\begin{equation}
\Sigma = \left(\dfrac{\beta^2}{k_P} - \dfrac{k_\phi^2}{2c} \right) \tilde{\psi}^2.
\end{equation}
It is worth noticing that, from one side, the liquid crystal network makes the lamellar structure stiffer, since the compression modulus is strengthened with respect to the polarization-free model, while on the other side, it counterbalances the negative surface tension of the lamellar phase.
\emph{In ultimis}, the coefficient of the laplacian term is half of the curvature modulus, which gives the energetic cost associated to an infinitesimal bending of a layer. It can be written as
\begin{equation}
\Upsilon = \dfrac{k_\phi}{2} \tilde{\psi}^2.
\end{equation}

\section{Polarization flip}\label{sec:polappendix}
In this Section we illustrate the mechanism at the origin of the flip of polarization at the boundary, giving rise to intermittent viscosity behaviour, characterizing the blue region of the phase diagram presented in Fig.~\ref{fig:phase_diagram}. 
This phenomenon is driven by the collision of active domains against the boundary layers. In Fig.~\ref{fig:S2} we show, for the case at $Er_{act}=0.57, Er=0.003$, a series of snapshots of a change of polarization on the bottom wall from the antiparallel alignment of $\vec{P}$ with respect to the imposed velocity ($-$ state) to a parallel configuration ($+$ state). The overall dynamics of the event can be better appreciated by looking at the attached Movie~3,4. 
Panel~(a) of Fig.~\ref{fig:S2} shows the configuration at time $t=52380$, before the event.
Panel~(b) shows a zoom at the same time of a colliding active domain, characterized by a typical vortical pattern in the polarization field, with a $+1$ defect, highlighted by a red square in panel~(b).
As the droplet marges with the layer, strong elastic interactions produce bendings of the liquid crystal network (see panel~(c) at time $t=53580$).  
In panel~(d) at a close subsequent time $t=56580$, the system is still found in a homogeneous $-$ $P_w$ state on the wall, despite the polarization on the top part of the layer is now directed in the same direction of the imposed flow. The $+1$ defect has disappeared due to the interaction with the wall, thus generating a complex rearrangement dynamics. 
As a result of the advection and the elastic deformations, the polarization flap, directed along the flow direction and highlighted by the black box in panel~(d), is pushed on the walls and adheres on it, resulting in a local change of the polarization state (see panel~(e) at time $t=58286$). 
After a similar event, not shown in Fig.~\ref{fig:S2}, the overall hydrodynamic state changes, thus leading to a final homogeneous $+$ state, shown in panel~(f).
We observe that flip dynamics generally takes place on time-scale of order $\sim\mathcal{O}(10^5)$ timesteps, much shorter than the lifetime of viscosity states that are found to be $\sim \mathcal{O}(10^7)$.

\end{document}